&pdflatex

\documentclass[11pt,prd,aps,amssymb,amsmath,tightenlines,showpacs]{revtex4}
\usepackage{graphicx}

\usepackage[T1]{fontenc}
\usepackage[latin1]{inputenc}
\usepackage[english]{babel}
\usepackage{amsmath}
\usepackage{amsfonts}
\usepackage{amssymb}

\def\half{{\textstyle{\frac{1}{2}}}}
\def\cP{\mathcal P}
\def\cT{\mathcal T}
\def\cH{\mathcal H}
\def\cPT{\mathcal{PT}}

\begin{document}
\title{Coupled Oscillator Systems Having Partial $\mathcal{PT}$ Symmetry}

\author{Alireza Beygi$^1$}\email{beygi@stud.uni-heidelberg.de}
\author{S. P. Klevansky$^{1,2}$}\email{spk@physik.uni-heidelberg.de}
\author{Carl M. Bender$^3$}\email{cmb@wustl.edu}

\affiliation{$^1$Institut f\"{u}r Theoretische Physik, Universit\"{a}t
Heidelberg, Philosophenweg 12, 69120 Heidelberg, Germany\\
$^2$Department of Physics, University of the Witwatersrand, Johannesburg,
South Africa\\
$^3$Department of Physics, Washington University, St. Louis,
Missouri 63130, USA} 

\begin{abstract}
This paper examines chains of $N$ coupled harmonic oscillators. In isolation,
the $j$th oscillator ($1\leq j\leq N$) has the natural frequency $\omega_j$ and is
described by the Hamiltonian $\half p_j^2+\half\omega_j^2x_j^2$. The oscillators
are coupled adjacently with coupling constants that are purely imaginary; the
coupling of the $j$th oscillator to the $(j+1)$st oscillator has the bilinear
form $i\gamma x_jx_{j+1}$ ($\gamma$ real). The complex Hamiltonians for these
systems exhibit {\it partial} $\cPT$ symmetry; that is, they are invariant under
$i\to-i$ (time reversal), $x_j\to-x_j$ ($j$ odd), and $x_j\to x_j$ ($j$ even).
[They are also invariant under $i\to-i$, $x_j\to x_j$ ($j$ odd), and $x_j\to-
x_j$ ($j$ even).] For all $N$ the quantum energy levels of these systems are
calculated exactly and it is shown that the ground-state energy is real. When
$\omega_j=1$ for all $j$, the full spectrum consists of a real energy spectrum
embedded in a complex one; the eigenfunctions corresponding to real energy
levels exhibit partial $\cPT$ symmetry. However, if the $\omega_j$ are allowed
to vary away from unity, one can induce a phase transition at which {\it all}
energies become real. For the special case $N=2$, when the spectrum is real, the
associated classical system has localized, almost-periodic orbits in phase space
and the
classical particle is confined in the complex-coordinate plane. However, when
the spectrum of the quantum system is partially real, the corresponding
classical system displays only open trajectories for which the classical
particle spirals off to infinity. Similar behavior is observed when $N>2$.
\end{abstract}
\maketitle

\section{Introduction}
\label{s1}
There are many experimental and theoretical studies of $\cPT$-symmetric
coupled-oscillator Hamiltonians \cite{r1,r2,r3,r4,r5,r6}. In most cases the
starting
point is either a coupled set of $\cPT$-symmetric equations of motion, or a
$\cPT$-symmetric Hamiltonian that governs such a system. It has been established
that such $\cPT$-symmetric systems exhibit a rich phase structure with phase
boundaries depending on the number of oscillators, how they are coupled, and the
values of the coupling parameters \cite{r5}.

In a recent paper on radiative coupling and weak lasing of exciton-polariton
condensates, Aleiner {\it et al.} \cite{r7} considered a Hamiltonian function
that governs condensation centers that are bilinearly coupled by a term of the
form $igzz^*$, where
each center is described by the complex coordinate $z$ and $g$ is a coupling
strength. They investigated the classical dynamics of the system.
While there is no obvious underlying symmetry, the authors found closed paths in
their spin trajectories. This intriguing result motives the current study of an
unusual type of oscillator system; namely, a chain of $N$ harmonic oscillators
with {\it pure imaginary coupling}. The Hamiltonian for the $j$th oscillator ($1
\leq j\leq N$) has the form $\half p_j^2+\half\omega_j^2x_j^2$, where the
natural
frequency $\omega_j$ is real and positive. The $j$th oscillator is coupled to
the $(j+1)$st oscillator by an imaginary coupling constant $i\gamma$, where
$\gamma$ is real and independent of $j$. The coupling term is {\it bilinear};
that is, it has the form $i\gamma x_jx_{j+1}$. The Hamiltonian that governs this
system of $N$ adjacently coupled oscillators has the form
\begin{equation}
H_N=\frac{1}{2}\sum_{j=1}^N\left(p_j^2+\omega_j^2x_j^2\right)+i\gamma
\sum_{j=1}^{N-1}x_jx_{j+1}\quad(N\geq2).
\label{E1}
\end{equation}

This complex Hamiltonian is not $\cPT$ symmetric because $i$ changes sign under
time reversal $\cT$ and it is assumed that every coordinate $x_j$ changes sign
under
parity $\cP$. However, $H_N$ is {\it partially} $\cPT$ symmetric; that is, it
remains invariant if we change the sign of $i$ and simultaneously reverse the
sign of only the odd-numbered or only the even-numbered coordinates. To
illustrate, we define $\cP_j$ as the operator that reverses the sign of $x_j$
but does not affect any other coordinate. Then, $H_2$ is partially 
$\cPT$ symmetric with respect to $\cP_1\cT$ and also with respect to $\cP_2\cT$.
Similarly, $H_3$ is partially $\cPT$ symmetric with respect to $\cP_1\cP_3\cT$
and also with respect to $\cP_2\cT$. Note that reversing the signs of an even
number of coordinates is achievable by a rotation but reversing the signs of an
odd number of coordinates is not achievable by a rotation. For example, for $N=
2$, $x_1\to-x_1$, $x_2\to-x_2$ is merely a rotation by an angle of $\pi$ in the
$x_1,x_2$ plane, but $x_1\to-x_1$, $x_2\to x_2$ cannot be achieved by a
rotation. For $N=3$, $\cP_1\cP_3$ is a rotation but $\cP_2$ and also $\cP_1\cP_2
\cP_3$ are not.

Systems having partial $\cPT$ symmetry have remarkable properties. In
Sec.~\ref{s2} we set $\omega_j=1$ for all $j$ and show that for small
$N$ and for all values of the coupling parameter $\gamma$ the ground-state
energy of the quantum system is real and positive. Then, in Sec.~\ref{s3} we
present the exact solution for the complete quantum spectrum for all $N$. We
find that the ground-state energy is always real, but that the full spectrum is
partly real and partly complex. For each energy, we calculate the corresponding
eigenfunction and demonstrate that simultaneous
eigenfunctions of the Hamiltonian and the partial $\cPT$ operator have real
energies, while those that are not partially $\cPT$ symmetric are associated
with complex energies. Thus, partial $\cPT$ symmetry is associated with a
partially real energy spectrum. In Sec.~\ref{s4} we relax the constraint that
$\omega_j=1$. We show that for $N=2$ it is possible to choose the natural
oscillator frequencies to make the energy spectrum completely real. Thus, there
is a phase transition from a partially real to a completely real spectrum. This
result is shown to hold in a modified form for $N=3$ and $N=4$. Next, in
Sec.~\ref{s5}, we investigate the classical solutions for the $N=2$ and $N=3$
systems and find no remnant of the partially $\cPT$-symmetric phase; that is,
all classical orbits are open unless the quantum spectrum is entirely real,
in which case the orbits are all closed and periodic. Brief concluding remarks
are given in Sec.~\ref{s6}.

\section{Ground-State Energies of $N$ Coupled Oscillators with $\omega_j=1$}
\label{s2}
In this section we show that the ground-state energy of a quantum system of $N$
coupled oscillators with natural frequency $\omega_j=1$ is real and positive.

\subsection{Two coupled oscillators}
Let us consider the quantum-mechanical Hamiltonian of two coupled oscillators
$H_2=\half p^2+\half q^2+\half x^2+\half y^2+i\gamma xy$, where $x$ and $y$ are
the coordinates, $p$ and $q$ are the conjugate momenta, $\gamma$ is a coupling
strength, and $\omega_1=\omega_2=1$. This Hamiltonian is partially 
$\cPT$ symmetric because it is invariant under the transformations $\cP_x\cT$ or
$\cP_y\cT$, where $\cP_x:\,(x,y)\to(-x,y)$, $\cP_y:\,(x,y)\to(x,-y)$, and
$\cT:\,i\to-i$. The Schr\"odinger equation associated with $H_2$ is 
\begin{equation}
\left(-\half\partial_x^2-\half\partial_y^2+\half x^2+\half y^2+i\gamma xy\right)
\psi(x,y)=E\psi(x,y).
\label{E2}
\end{equation}
The ground-state eigenfunction has the (non-nodal) gaussian form
\begin{equation}
\psi_0(x,y)=\exp\left(-\half ax^2-\half ay^2+bxy\right),
\label{E3}
\end{equation}
where $a$ and $b$ are constants. Note that $\psi_0(x,y)$ is $\cPT$ symmetric in
either $x$ or $y$. Inserting (\ref{E3}) into (\ref{E2}) and matching powers of
$x$ and $y$ gives the three equations $E_0=a$, $a^2+b^2=1$, and $2ab=-i\gamma$.

The physically acceptable solution to these equations requires that $b$ be
imaginary, $b=-i\frac{\gamma}{2a}$, and that $E_0=a$ be the real and positive
solution to $a^4-a^2-\gamma^2/4=0$,
\begin{equation}
E_0=a=\left(\half+\half\sqrt{1+\gamma^2}\right)^{1/2}.
\label{E4}
\end{equation}
Note that because $b$ is imaginary and $a$ is real and positive, $\psi_0(x,y)$
vanishes as $x^2+y^2\to\infty$.

\subsection{Three coupled oscillators}
For three oscillators the Hamiltonian $H_3$ in (\ref{E1}) with $\omega_j=1$ has
the form
\begin{equation}
H_3=\half p^2+\half q^2+\half r^2+\half x^2+\half y^2+\half z^2+i\gamma(xy+yz).
\label{E5}
\end{equation}
Again, $H_3$ is partially $\cPT$ symmetric; it is invariant under $\cP_y\cT$
(and also $\cP_x\cP_z\cT$). The lowest-energy eigenstate has the form
$\psi_0(x,y,z)=\exp\left[-\half a(x^2+z^2)-\half by^2+c(xy+yz)+dxz\right],$
where $a$, $b$, $c$, and $d$ are constants. Solving the Schr\"odinger equation
$H_3\psi_0(x,y,z)=E\psi_0(x,y,z)$ and comparing powers in $x$, $y$, and $z$
gives the five equations $E_0=a+\half b$, $1=a^2+d^2+c^2$, $1=2c^2+b^2$, $i
\gamma=c(d-a-b)$, $2ad=c^2$. We solve these equations and verify that the
eigenfunction is normalizable [$\psi_0(x,y,z)$ vanishes as $x^2+y^2+z^2\to
\infty$] and that, even though $H_3$ is complex, the ground-state energy is
real and positive,
\begin{equation}
E_0=\half+\half\left(2+2\sqrt{1+2\gamma^2}\right)^{1/2}.
\label{E6}
\end{equation}
The ground-state eigenfunction $\psi_0(x,y,z)$ has partial $\cP\cT$-symmetry.
Also, in the limit $\gamma\to0$ the oscillators decouple and we recover the
expected result that $E_0=3/2$.

\subsection{Four coupled oscillators}
For four coupled oscillators the coordinates are $x,y,z,w$, the
canonical momenta are $p,q,r,s$, the Hamiltonian $H_4$ with $\omega_j=1$ is
partially $\cPT$ symmetric in the variables $x,z$ or $y,w$, and reads 
\begin{equation}
H_4=\half\left(p^2+q^2+r^2+s^2\right)+\half\left(x^2+y^2+z^2+w^2\right)+i\gamma
(xy+yz+zw).
\label{E7}
\end{equation}
We solve the Schr\"odinger equation $H_4\psi_0=E\psi_0$ with the {\it ansatz}
for a partially $\cPT$-symmetric ground-state wave function of gaussian form
\begin{equation}
\psi_0(x,y,z,w)=\exp\left[-\frac{a}{2}x^2-\frac{b}{2}y^2-\frac{b}{2}z^2-
\frac{a}{2}w^2+c(xy+zw)+d(xz+yw)+exw+fyz\right],
\label{E8}
\end{equation}
where $a,b,c,d,e,f$ are six arbitrary constants. This leads to the conditions
$E_0=a+b$, $f^2+d^2+c^2+b^2=1$, $e^2+d^2+c^2+a^2=1$, $i\gamma=2cd-2bf$, $cf+ce=
bd+ad$, $i\gamma=df+de-cb-ca$, $ae=cd$. Clearly, the complexity of
the coupled nonlinear system of equations increases rapidly as the number of
coupled oscillators increases.

For the case of $N$ coupled oscillators with $\omega_j=1$, we show in
Sec.~\ref{s3} that the ground-state energy $E_0$ is
\begin{equation}
E_0=\frac{1}{2}\sum_{j=1}^N\sqrt{1+2i\gamma\cos\left[j\pi/(N+1)\right]}.
\label{E10}
\end{equation}
By setting $N=2$ or $N=3$, we readily recover (\ref{E4}) and
(\ref{E6}). For $N=4$, (\ref{E10}) yields the value
\begin{equation}
E_0=\left[\half+\half\sqrt{1+\gamma^2(3+\sqrt{5})/2}\right]^{1/2}+
\left[\half+\half\sqrt{1+\gamma^2(3-\sqrt{5})/2}\right]^{1/2}.
\label{E11}
\end{equation}
Closer inspection of (\ref{E10}) reveals that the ground-state energy of such
coupled oscillators is always real. Indeed, (\ref{E10}) can be rewritten as
\begin{equation}
E_0=\frac{1}{2}\sum_{j=1}^N\left(\half+\half\sqrt{1+4\gamma^2\cos^2\left[
j\pi/(N+1)\right]}\right)^{1/2}.
\label{E12}
\end{equation}

\section{Exact eigenfunctions and spectra of $N$ coupled oscillators}
\label{s3}

\subsection{Two coupled oscillators}
Let us return to the two-coupled-oscillator system governed by the Hamiltonian
$H_2$ with $\omega_j=1$. The transformation $x_1=(x+y)/\sqrt{2}$, $x_2=(x-y)/
\sqrt{2}$ decouples the oscillators, leading to the Hamiltonian $H=\half p_1^2+
\half(1+i\gamma)x_1^2+\half p_2^2+\half(1-i\gamma)x_2^2$, which has
complex-conjugate frequencies $\nu_1^2=1+i\gamma$ and $\nu_2^2=1-i\gamma$.
Apart from a normalization constant, the eigenfunctions are
\begin{equation} 
\Psi_{n_1,n_2}(x_1,x_2)=\cH_{n_1}(\sqrt{\nu_1}x_1)\cH_{n_2}(\sqrt{\nu_2}x_2)
e^{-\nu_1 x_1^2/2}e^{-\nu_2 x_2^2/2}
\label{E13}
\end{equation}
with corresponding energy eigenvalues $E_{n_1,n_2}=\nu_1\left(n_1+\half\right)+
\nu_2\left(n_2+\half\right)$. In terms of the coupling parameter $\gamma$ the
frequencies are
$$\nu_1=\nu_2^*=\left(\half+\half\sqrt{1+\gamma^2}\right)^{1/2}+i
\left(-\half+\half\sqrt{1+\gamma^2}\right)^{1/2},$$
whose real parts are positive. The general result for the energy spectrum is
$$E_{n_1,n_2}=\left(\half+\half\sqrt{1+\gamma^2}\right)^{1/2}(n_1+n_2+1) 
+i\left(-\half+\half\sqrt{1+\gamma^2}\right)^{1/2}(n_1-n_2).$$

\begin{figure}[h]
\begin{center}
\includegraphics[scale=0.60]{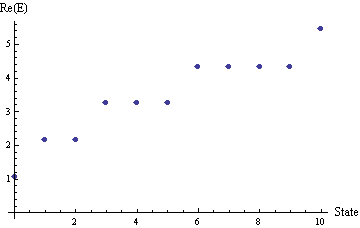}
\hspace{0.3cm}
\includegraphics[scale=0.60]{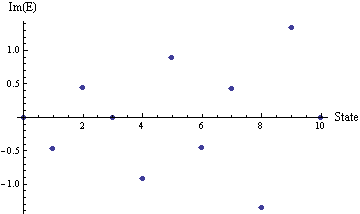}
\end{center}
\caption{Real parts (left panel) and imaginary parts (right panel) of the first
eleven energy levels of $H_2$ ($N=2$, $\gamma=1$).}
\label{F1} 
\end{figure}

Note that the spectrum is real if $n_1=n_2$. If $n_1=n_2=0$, we recover the
ground-state energy in (\ref{E4}). In addition, we obtain the corresponding
eigenfunction from (\ref{E13}),
$$\Psi_{0,0}(x,y)=\exp\left[-\half\left(\half+\half\sqrt{1+\gamma^2}
\right)^{1/2}(x^2+y^2)-i\left(-\half+\half\sqrt{1+\gamma^2}\right)^{1/2}xy
\right],$$
which verifies the {\it ansatz} (\ref{E3}) and explicitly demonstrates that an
eigenfunction having the partial $\cPT$ symmetry of the Hamiltonian is
associated with a real eigenvalue. Note also that the real spectrum is part of a
larger spectrum containing complex-conjugate pairs. This can be illustrated by
the choice $n_1=1$ and $n_2=0$ or $n_1=0$ and $n_2=1$:
$$E_{1,0}=\left(2+2\sqrt{1+\gamma^2}\right)^{1/2}+i\left(-\half+\half\sqrt{1+
\gamma^2}\right)^{1/2}$$
and $\Psi_{1,0}(x,y)=\sqrt{2}(1+i\gamma)^{1/4}(x+y)\Psi_{0,0}$, which is neither
$\cP_x\cT$ nor $\cP_y\cT$-symmetric. In addition, $E_{0,1}=E_{1,0}^*$ and
$\Psi_{0,1}(x,y)=\Psi_{1,0}(x,y)^*$. The real parts of the energies are $n_1+n_2
+1$-fold degenerate, as shown in Fig.~\ref{F1}.

\begin{figure*}
\includegraphics[width=0.8\textwidth]{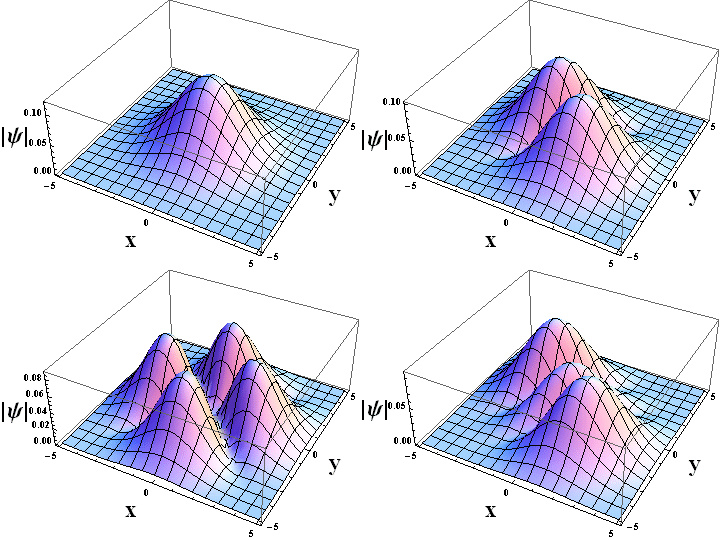}
\caption{(Color online) Absolute value of the eigenfunction for the ground
state (upper left), $n_1=n_2=1$ (lower left), $n_1=0,\,n_2=1$ (upper right) and 
$n_1=0,\,n_2=2$ (lower right) for $\gamma=1$.}
\label{F2} 
\end{figure*}

The nature of the eigenfunctions associated with the first few energy levels is
depicted in Fig.~\ref{F2}. The ground-state ($n_1=n_2=0$) and the
third-exited-state ($n_1=n_2=1$) eigenfunctions are partially
$\cPT$ symmetric, as can be seen in the left upper and lower diagrams, while the
eigenfunctions corresponding to the first and third (complex) eigenvalues ($n_1=
0,\,n_2=1$ and $n_1=0,\,n_2=2$), which are not partially $\cPT$ symmetric are
shown on the right diagrams.

\subsection{Three coupled oscillators}
\begin{figure}[h]
\begin{center}
\includegraphics[scale=0.60]{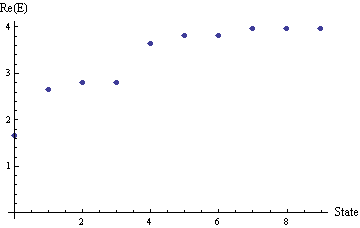}
\hspace{0.3cm}
\includegraphics[scale=0.60]{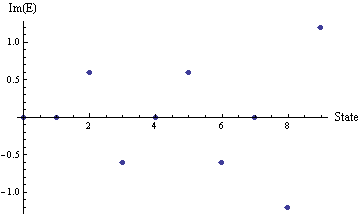}
\end{center}
\caption{Real parts (left panel) and imaginary parts (right panel) of the
first ten energies for $H_3$ ($\gamma=1$, $N=3$).}
\label{F3} 
\end{figure}

To find the exact solution to the Schr\"odinger equation for $H_3$ in (\ref{E5})
with $\omega_j=1$ we make the transformation $x_1=(x-z)/\sqrt{2}$, $x_2=y/
\sqrt{2}+(x+z)/2$, $x_3=-y/\sqrt{2}+(x+z)/2$, which decouples the oscillators,
giving $H=\half p_1^2+\half\nu_1^2x_1^2+\half p_2^2+\half\nu_2^2x_2^2+\half
p_3^2+\half\nu_3^2x_3^2$ with $\nu_1^2=1$, $\nu_2^2=1+i\gamma\sqrt{2}$, and
$\nu_3^2=1-i\gamma\sqrt{2}$. Thus, the unnormalized eigenfunctions are
$$\Psi(x_1,x_2,x_3)=\cH_{n_1}(x_1)\cH_{n_2}(\sqrt{\nu_2}x_2)\cH_{n_3}(\sqrt{
\nu_3}x_3)\exp\left[-(x_1^2 +\nu_2x_2^2+\nu_3x_3^2)/2\right]$$
and the energies are $E=n_1+\half+\nu_2\left(n_2+\half\right)+\nu_3\left(n_3+
\half\right)$, where
$$\nu_{2,3}=\left(\half+\half\sqrt{1+2\gamma^2}\right)^{1/2}\pm i\left(-\half+
\half\sqrt{1+2\gamma^2}\right)^{1/2}.$$
Thus, the energy spectrum can be expressed as
$$E=n_1+\half+\sqrt{\half+\half\sqrt{1+2\gamma^2}}+
\sqrt{\half+\half\sqrt{1+2\gamma^2}}\,\left(n_2+n_3\right) 
+i\sqrt{-\half+\half\sqrt{1+2\gamma^2}}\,\left(n_2-n_3\right).$$
Evidently, if the second and third oscillators are in the same state ($n_2=
n_3$), the energy is real and the corresponding eigenfunctions are partially
$\cPT$ symmetric. In particular, the ground-state energy (\ref{E6}) is recovered
and the ground-state eigenfunction is
$$\Psi_{0,0,0}(x,y,z)=\exp\left[-(1+a)(x^2+z^2)/4-ay^2/2-(a-1)xz/2-ib(xy+yz)/
\sqrt{2}\right],$$
where $a={\rm Re}\,\nu_2$, $b={\rm Im}\,\nu_2$. The first ten energies of
$H_3$ for $\gamma=1$ are shown in Fig.~\ref{F3}.

\subsection{Four coupled oscillators}
For the Hamiltonian (\ref{E7}), which governs four linearly coupled oscillators,
the transformation
\begin{eqnarray}
x_{1,2}&=&\frac{1}{2\sqrt{5}}\left[\sqrt{5-\sqrt{5}}\,(x\pm w)
\pm\sqrt{5+\sqrt{5}}\,(y\pm z)\right],\nonumber\\
x_{3,4}&=&\frac{1}{2\sqrt{5}}\left[\sqrt{5+\sqrt{5}}\,(x\mp w)
\pm\sqrt{5-\sqrt{5}}\,(y\mp z)\right]\nonumber
\end{eqnarray}
exactly decouples the oscillators. The new Hamiltonian takes the form
$$H=\half p_1^2+\half\nu_1^2x_1^2+\half p_2^2+\half\nu_2^2x_2^2+\half p_3^2+
\half\nu_3^2x_3^2+\half p_4^2+\half\nu_4^2x_4^2,$$
where $\nu_1^2=\nu_2^{2*}=1+\half i\gamma(1+\sqrt5)$ and $\nu_3^2=\nu_4^{2*}=1+
\half i\gamma(-1+\sqrt5)$ are complex frequencies. Let ${\rm Re}\,\nu_1=
{\rm Re}\,\nu_2=A$, ${\rm Re}\,\nu_3={\rm Re}\,\nu_4=C$, ${\rm Im}\,\nu_1=-{\rm
Im}\,\nu_2=B$, ${\rm Im}\,\nu_3=-{\rm Im}\,\nu_4=D$, where
$$(A,B)=\sqrt{\pm\half+\half\sqrt{1+\half\gamma^2(3+\sqrt{5})}},\quad
(C,D)=\sqrt{\pm\half+\half\sqrt{1+\half\gamma^2(3-\sqrt{5})}}.$$
In terms of these variables and the quantum numbers $n_1$, $n_2$, $n_3$,
$n_4$, the total energy is
$$E_{n_1,n_2,n_3,n_4}=A(n_1+n_2+1)+C(n_3+n_4+1)+iB(n_1-n_2)+iD(n_3-n_4).$$
Thus, if $n_1=n_2$ and $n_3=n_4$, the energy is real. When the energy is real,
the corresponding eigenfunction is always $\cP_{xz}\cT$- or $\cP_{yw}\cT$-
symmetric. For example, the ground-state energy is $E_{0,0,0,0}=A+C$ and the
corresponding eigenfunction is
\begin{eqnarray}
&&\Psi_{0,0,0,0}=\exp\left[-\frac{1}{4}\left(A+C+\frac{C-A}{\sqrt{5}}\right)
\left(x^2+w^2\right)-\frac{1}{4}\left(A+C+\frac{A-C}{\sqrt{5}}\right)
\left(y^2+z^2\right)\right.
\nonumber\\
&&-\frac{1}{\sqrt{5}}(A-C)(xz+yw)-\frac{i}{2}\left(B-\frac{B}{\sqrt{5}}
-D-\frac{D}{\sqrt{5}}\right)xw-\frac{i}{2}\left(B+
\frac{B}{\sqrt{5}}-D+\frac{D}{\sqrt{5}}\right)yz\nonumber\\
&&-\left.\frac{i}{\sqrt{5}}(B+D)(xy+wz)\right].
\end{eqnarray}
This eigenfunction displays the symmetries assumed in the {\it ansatz}
(\ref{E8}).

As another illustration, we consider the case in which the first two oscillators
are in the first excited state, and the other two in the ground state
($n_1=n_2=1$, $n_3=n_4=0$). The energy is $E_{1,1,0,0}=3A+C$ and the
eigenfunction is
$$\Psi_{1,1,0,0}=\frac 15\sqrt{A^2+B^2}\left[(5-\sqrt 5)(x^2-\omega^2)-
(5+\sqrt{5})(y^2-z^2)+4\sqrt{5}xz-4\sqrt{5}yw\right]\Psi_{0,0,0,0}.$$
Once again, the energy is real and the eigenfunction is $\cP_{xz}\cT$- or
$\cP_{yw}\cT$-symmetric. A complex energy $E_{1,0,0,0}=2A+C+iB$ arises for the
choice $n_1=1$, $n_2=n_3=n_4=0$.

\subsection{Five coupled oscillators}
The Hamiltonian for five coupled oscillators is
$$H_5=\half\left(p^2+q^2+r^2+s^2+t^2\right)+\half\left(x^2+y^2+z^2
+w^2+u^2\right)+i\gamma(xy+yz+zw+wu).$$
Rather than decoupling the oscillators we treat this case by
constructing the secular equation 
\begin{equation}
\label{E15}
{\rm det}\left(M_{jk}-\nu^2\delta_{jk}\right)=0,
\end{equation}
where $M_{jk}$ is the tridiagonal matrix defined as
\begin{equation}
\label{E16}
M_{jk}\equiv\frac{\partial^2 U}{\partial q_j\partial q_k}\bigg|_0=
\left(\begin{array}{ccccc}
1 & i \gamma & 0 & 0 & 0 \\ i \gamma & 1 & i \gamma & 0 & 0 \\
0 & i \gamma & 1 & i \gamma & 0 \\ 0 & 0 & i \gamma & 1 & i \gamma \\
0 & 0 & 0 & i \gamma & 1\end{array}\right),
\end{equation}
$U$ is the potential, and $q_j$ and $q_k$ are coordinates.

The solution to the secular equation (\ref{E15}) gives complex-conjugate pairs
of frequencies and one real frequency: $\nu_{1,2}^2={1\pm i\gamma\sqrt{3}}$,
$\nu_{3,4}^2={1\pm i\gamma}$, $\nu_5=1$. Thus, the decoupled Hamiltonian is
$$H=\half p_1^2+\half\nu_1^2 x_1^2+\half p_2^2+\half\nu_2^2 x_2^2+\half p_3^2+
\half\nu_3^2 x_3^2+\half p_4^2+\half\nu_4^2 x_4^2+\half p_5^2+\half x_5^2$$
and the energy of the system reads
$$E=\nu_1\left(n_1+\half\right)+\nu_2\left(n_2+\half\right)+\nu_3\left(n_3+\half
\right)+\nu_4\left(n_4+\half\right)+n_5+\half,$$
where $n_1$, $n_2$, $n_3$, $n_4$, $n_5$ are nonnegative integers.

\subsection{General case: $N$ coupled oscillators with $\omega_j=1$}
In this section we consider the Hamiltonian (\ref{E1}) for $N$ linearly coupled
oscillators with $\omega_j=1$.
To obtain the frequencies of the decoupled oscillators we use (\ref{E15}) to
construct the $N\times N$ tridiagonal matrix secular equation, ${\rm det}(
\mathbf{M}-\nu^2\mathbf{I})=0$, which has the form
\begin{equation}
D_N=\begin{vmatrix}
1 - \nu^2 & i \gamma &  &  &  &  \\
i \gamma & 1 - \nu^2 & i \gamma &  &  &  \\
 & i \gamma & 1 - \nu^2 & i \gamma &  &  \\
 &  & \ddots & \ddots & \ddots &  \\
 &  &  & i \gamma & 1 - \nu^2 & i \gamma \\
 &  &  &  & i \gamma & 1 - \nu^2 \end{vmatrix}=0.\nonumber
\end{equation}
Because this matrix equation is tridiagonal, $D_N$ satisfies the three-term
recurrence relation
$$D_{k} +\left(\nu^2-1\right)D_{k-1}-\gamma^2 D_{k-2}=0\quad(k=1,2,\ldots,N)
,$$
where $D_0=1$ and $D_{-1}=0$. We solve this difference equation to obtain the
frequencies
$$\nu^2=1+2i\gamma\cos[j\pi/(N+1)]\quad(j=1,2,\ldots,N).$$
Thus, the exact expression for the total energy of the system of $N$ oscillators
is given by
$$E=\sum_{j=1}^N\sqrt{1+2i\gamma\cos[j\pi/(N+1)]}\,\left(n_j+\half\right),$$
where $n_j\geq0$ ($j=1,...,N$). Choosing $n_j=0$ for all $j$, we find the exact
ground-state energy in (\ref{E12}), which has been shown to be real.

\section{Coupled oscillators with arbitrary frequencies}
\label{s4}
In Secs.~\ref{s2} and \ref{s3} the oscillator frequencies multiplying $x_j^2$
were set to unity. Our conclusion in the foregoing analysis was that a real
spectrum is embedded in a complex spectrum containing complex-conjugate pairs of
energies. We now relax this constraint on the natural frequencies. For the
two-, three-, and four-coupled-oscillator systems, we demonstrate that for
an appropriate choice of $\omega_j$ the spectrum can be entirely real.

\subsection{Two coupled oscillators with general natural frequencies
$\omega_x$ and $\omega_y$}
The Hamiltonian $H_2$ in (\ref{E1}) reads 
$H_2=\half p^2+\half q^2+\half\omega_x^2x^2+\half\omega_y^2y^2+i\gamma xy$.
The frequencies of the decoupled oscillators in this case are
\begin{equation}
\nu_{1,2}^2=\half\left({\omega_x^2+\omega_y^2\pm\sqrt{(\omega_x^2-\omega_y^2)^2
-4\gamma^2}}\right)
\label{E19}
\end{equation}
and the energies of the system are $E_{n_1,n_2}=\nu_1\left(n_1+\half\right)
+\nu_2\left(n_2+\half\right)$, where $n_1,n_2\geq0$.

In contrast to the results found in Sec.~\ref{s3}, the entire energy
spectrum can be real for specific values of $\omega_x$, $\omega_y$, and
$\gamma$. For this to be so, the parameters must satisfy the condition
\begin{equation}
\label{E20}
|\omega_x^2-\omega_y^2|\geq2|\gamma|.
\end{equation}
The case considered in Sec.~\ref{s3} had $\omega_x=\omega_y=1$ and
$\gamma=1$, which does not satisfy this condition and, as we saw, the energy
spectrum was only partially real.

To have real eigenvalues the associated eigenfunctions must all be partially
$\cPT$ symmetric. This can be seen explicitly by decoupling the oscillators with
the transformation
$$x_1=\sqrt{A+B}(Dx+Ex+iCy)/(2CE),\quad x_2=\sqrt{A-B}(-Dx+Ex-iCy)/(2CE),$$
where $A=8\gamma^2-2(w_x^2-w_y^2)^2$, $B=2(w_x^2-w_y^2)\big[(w_x^2-w_y^2)^2-4
\gamma^2\big]^{1/2}$, $C=2\gamma$, $D=w_x^2-w_y^2$, $E=\big[(w_x^2-w_y^2)^2-
4\gamma^2\big]^{1/2}$, leading to the Hamiltonian $H=\half p_1^2+\half\nu_1^2
x_1^2+\half p_2^2+\half\nu_2^2x_2^2$, where $\nu_1$ and $\nu_2$ are given in
(\ref{E19}). Up to a normalization constant, the eigenfunctions are
$$\Psi_{n_1,n_2}(x_1,x_2)=\cH_{n_1}\left(\sqrt{\nu_1}x_1\right)\cH_{n_2}
\left(\sqrt{\nu_2}x_2\right)\exp\left(-\half\nu_1x_1^2-\half\nu_2x_2^2\right).$$
Note that $A+B$ and $A-B$ have opposite signs in the $\cPT$-symmetric phase and
that $A-B<0$. Rewriting $\Psi_{n_1,n_2}(x_1,x_2)$ in terms of the original
variables $x$ and $y$, one can show that the eigenfunction $\Psi_{n_1,n_2}(x,y)$
has partial $\cPT$ symmetry because
$$(\cP_x\cT)\Psi_{n_1,n_2}(x,y)=(-1)^{n_1}\Psi_{n_1,n_2}(x,y),\quad
(\cP_y\cT)\Psi_{n_1,n_2}(x,y)=(-1)^{n_2}\Psi_{n_1,n_2}(x,y).$$

\begin{figure}[h]
\begin{center}
\includegraphics[scale=.7]{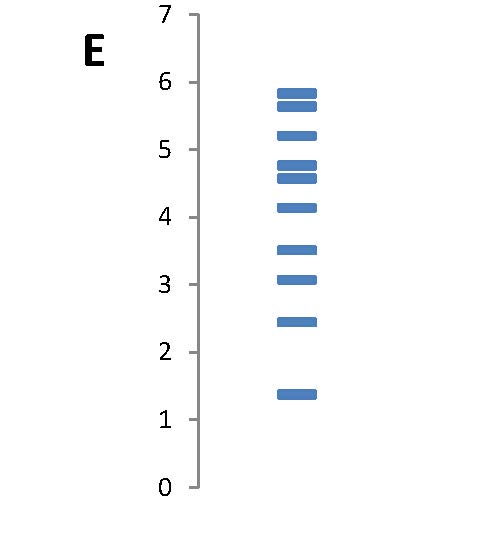}
\end{center}
\caption{First ten energies of $H_2$ for the parameter choice $\omega_x^2=
3$, $\omega_y^2=1$, and $\gamma=1/2$. These states have the quantum numbers
$(n_1,n_2)=(0,0),(0,1),(1,0),(0,2),(1,1),(0,3),(2,0),(1,2),(0,4),(2,1)$.}
\label{F4} 
\end{figure}

To illustrate, we consider the case $\omega_x^2=3$, $\omega_y^2=1$, and $\gamma=
1/2$. The relation (\ref{E20}) is satisfied and we find the purely real
nondegenerate spectrum shown in Fig.~\ref{F4}. When $|\omega_x^2-\omega_y^2|<2
|\gamma|$, the energy spectrum is only partially real. Thus, there is a phase
transition from the unbroken partially $\cPT$-symmetric phase to the broken one.
For example, keeping $\omega_y=1$ and $\gamma=1/2$, but adjusting $\omega_x^2$
so that it passes $2$, the first-excited-state energy becomes complex.

\subsection{Three coupled oscillators with general natural frequencies
$\omega_x$, $\omega_y$ and $\omega_z$}
\begin{figure}[!htb]
\centering
\includegraphics[width = 140mm]{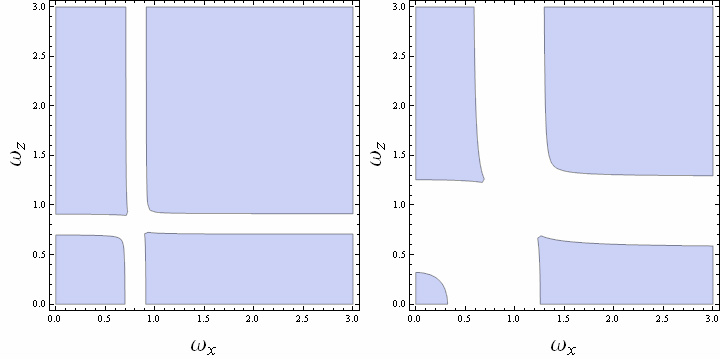}
\caption{(Color online) Unbroken partial $\mathcal{PT}$-symmetric phases of
$H_3$
depicted as shaded areas for the parameters $\gamma ={1}/{12}$ and $\omega_y^2
={2}/{3}$ (left panel) and $\gamma={1}/{3}$ and $\omega_y^2=1$ (right panel).}
\label{F5}
\end{figure}

For the three-oscillator Hamiltonian $H_3=\half p^2+\half q^2+\half
r^2+\half\omega_x^2x^2+\half\omega_y^2y^2+\half \omega_z^2z^2+i\gamma(xy+yz)$
the frequencies $\nu^2=\lambda$ of the decoupled oscillators satisfy the
cubic equation $f(\lambda)=0$, where
$$f(\lambda)=\lambda^3-\left(\omega_x^2+\omega_y^2+\omega_z^2\right)\lambda^2+
\left(\omega_x^2\omega_y^2+\omega_x^2\omega_z^2+\omega_y^2\omega_z^2+2\gamma^2
\right)\lambda-\omega_x^2\omega_y^2\omega_z^2-\left(\omega_x^2+\omega_z^2\right)
\gamma^2.$$
If the discriminant associated with this equation is positive, three real
distinct roots can emerge, giving a real spectrum. To guarantee that the roots
are positive, it is necessary that
$$f(0)<0,\quad \lambda_{\rm max}>0,\quad\lambda_{\rm min}>0,\quad f(\lambda_{\rm
max})>0,\quad f(\lambda_{\rm min})<0$$
are all fulfilled, with $\lambda_{\rm min}$ and $\lambda_{\rm max}$ being the
extrema of the polynomial. Figure~\ref{F5} displays the regions in which the
frequencies of the decoupled oscillators are all real (blue shaded areas) in the
parametric space of $\omega_x$ and $\omega_z$ for fixed values of $\gamma$
and $\omega_y^2$. This figure shows that several regions of unbroken
symmetry exist. This is in contrast to the case of the two coupled oscillators. 

\begin{figure}[!htb]
\centering
\includegraphics[width=70mm]{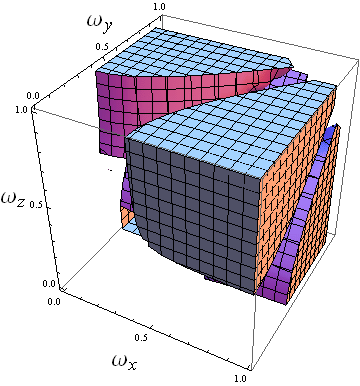}
\caption{(Color online) Unbroken partial $\cPT$-symmetric phases of $H_3$
for $\gamma=1/12$ depicted as colored volumes.}
\label{F6}
\end{figure}

For example, Figure~\ref{F5} shows that $\omega_x^2=1/3$, $\omega_y^2=2/3$,
$\omega_z^2=1$, and $\gamma=1/12$ gives an unbroken symmetry phase. We obtain
three different real positive (decoupled) frequencies 
$$\nu_1=\sqrt{2/3},\quad\nu_2=\sqrt{(8+\sqrt{14})/12},\quad\nu_3=\sqrt{(8-
\sqrt{14})/12}.$$
Thus, the spectrum is entirely real with energies given by $E_{n_1,n_2,n_3}=
\nu_1\left(n_1+\half\right)+\nu_2\left(n_2+\half\right)+\nu_3\left(n_3+\half
\right)$. By fixing only $\gamma$ we can find regions in the three-dimensional
parameter space of $\omega_x$, $\omega_y$, and $\omega_z$ for which unbroken
symmetry (and therefore a real spectrum) exists. This is shown in the colored
volumes depicted in Fig.~\ref{F6} for the specific choice $\gamma=1/12$. 

\subsection{Four coupled oscillators with general natural frequencies
$\omega_x$, $\omega_y$, $\omega_z$, and $\omega_w$}
The previous analysis can be applied to the four-coupled-oscillator Hamiltonian
$$H_4=\half p^2+\half q^2+\half r^2+\half s^2+\half\omega_x^2 x^2+\half
\omega_y^2 y^2+\half\omega_z^2 z^2+\half\omega_w^2 w^2+i\gamma(xy+yz+zw),$$
where $\omega_x$, $\omega_y$, $\omega_z$, and $\omega_w$ are real frequencies
and $\gamma$ is a real coupling parameter.

The eigenvalues of the matrix
$$M=\left(\begin{array}{cccc}
\omega_x^2 & i\gamma & 0 & 0 \\ i\gamma & \omega_y^2 & i\gamma & 0 \\ 0 &
i\gamma & \omega_z^2 & i\gamma\\ 0 & 0 & i\gamma &\omega_w^2\end{array}\right)$$
are the squares of the corresponding decoupled-oscillator frequencies. The
eigenvalues $\nu^2=\lambda$ satisfy the fourth-order equation
$f(\lambda)=\lambda^4-a\lambda^3+b\lambda^2-c\lambda+d=0$, where
\begin{align}
a &= \omega_x^2+\omega_y^2+\omega_z^2+\omega_w^2,\nonumber\\
b &= \omega_x^2\omega_y^2+\omega_x^2\omega_z^2+\omega_x^2\omega_w^2+\omega_y^2
\omega_z^2+\omega_y^2\omega_w^2+\omega_z^2\omega_w^2+3\gamma^2,\nonumber\\
c &= \omega_x^2\omega_y^2\omega_z^2+\omega_x^2\omega_y^2\omega_w^2+\omega_x^2
\omega_z^2\omega_w^2+\omega_y^2\omega_z^2\omega_w^2+2\gamma^2\omega_x^2+2
\gamma^2\omega_w^2+\gamma^2\omega_y^2+\gamma^2\omega_z^2,\nonumber\\
d &= \omega_x^2\omega_y^2\omega_z^2\omega_w^2+\gamma^2\omega_x^2\omega_y^2+
\gamma^2\omega_x^2\omega_w^2+\gamma^2\omega_z^2\omega_w^2+\gamma^4.\nonumber
\end{align}

\begin{figure}
\centering
\includegraphics[width=140mm]{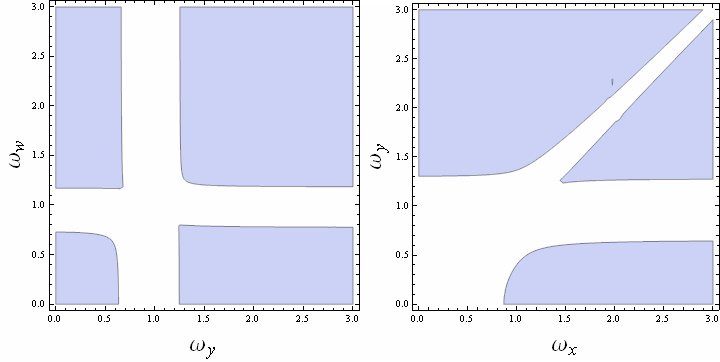}
\caption{(Color online) Unbroken partial $PT$-symmetric phases of $H_4$ depicted
as shaded area. Left panel: $\gamma=1/5$, $\omega_x^2=1$, and $\omega_z^2=1$.
Right panel: $\gamma=3/10$, $\omega_z^2=1$, and $\omega_w^2=4$.}
\label{F7}
\end{figure}

\begin{figure}
\centering
\includegraphics[width=140mm]{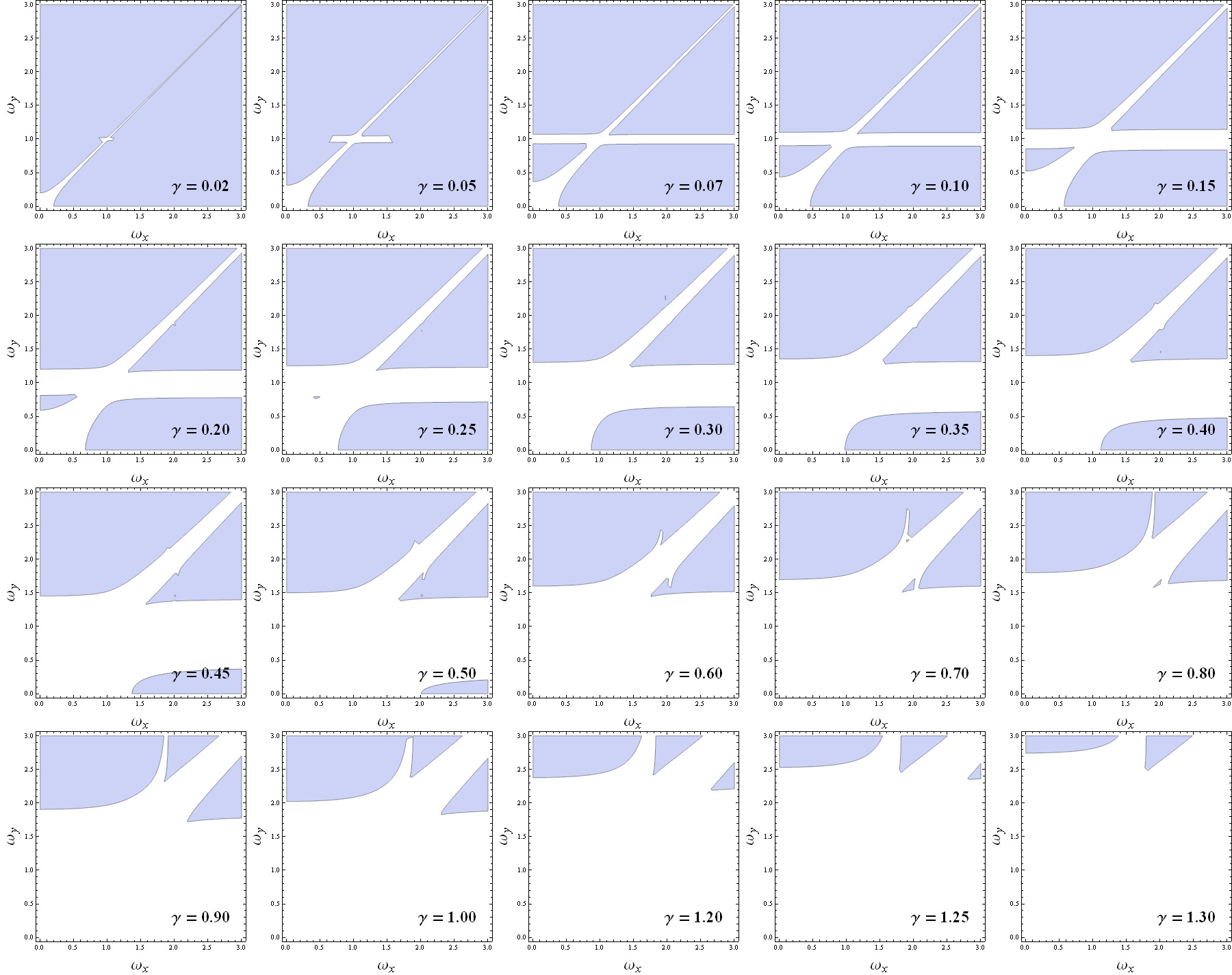}
\caption{(Color online) Unbroken partial $PT$-symmetric phases of $H_4$ in the
$(\omega_x,\omega_y)$ plane depicted as shaded areas for the parameter choices
$\omega_z^2=1$ and $\omega_w^2=4$ for twenty values of $\gamma$.}
\label{F75}
\end{figure}

Regions in which all decoupled oscillator frequencies are real give a
completely real energy spectrum, which means that partial $\cPT$ symmetry is
unbroken. This requires that $f(\lambda)$ have four positive roots, which is the
case if $f(0)>0$. In addition, if $f'(\lambda)$ has three positive roots,
the extrema of $f(\lambda)$ lie on the positive abscissa. To have four real
roots the minimum value of $f(\lambda)$ must be negative, and the maximum value
must be positive. Figure~\ref{F7} shows the regions in which these conditions
are fulfilled (that is, the regions in which the partial $\cPT$ symmetry is
unbroken) for specific choices of the parameters. Fixing the values of
$\omega_z^2=1$ and $\omega_w^2=4$ as in the right panel of Fig.~\ref{F7},
we can investigate the development of the phase boundaries as a function of the
coupling strength $\gamma$. This is shown in Fig.~\ref{F75}.

\section{Corresponding Classical Theory}
\label{s5}
\begin{figure}[h]
\begin{center}
\includegraphics[width = 140mm]{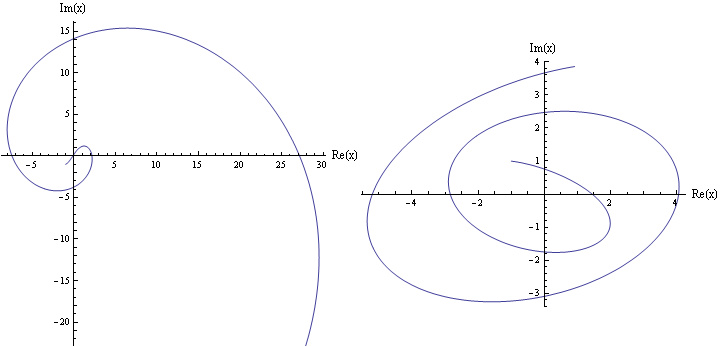}
\end{center}
\caption{Left panel: classical trajectory in the complex-$x$ plane for $H_2$
with $\omega_x^2=\omega_y^2=1$ with the initial conditions $x(0)=y(0)=-1-i$ and
$\dot{x}(0)=\dot{y}(0)=1+i/2$, and $\gamma=1$. Right panel: Classical trajectory
in the complex-$x$ plane for $H_2$ with $\omega_x^2=2$, $\omega_y^2=1$, and
$\gamma=1/2$ for the initial conditions $x(0)=y(0)=-1+i$ and $\dot{x}(0)=\dot{y}
(0)=2-i/4$.}
\label{F8}
\end{figure}

In this section we investigate the features of partially $\cPT$-symmetric
classical theories. We begin with the two-coupled-oscillator Hamiltonian $H_2$
with frequencies $\omega_x=\omega_y=1$. Hamilton's classical equations of motion
lead to
$$x''(t)+x(t)+i\gamma y(t)=0,\quad y''(t)+y(t)+i\gamma x(t)=0$$
and combining these equations gives the fourth-order differential equation
$$x''''(t)+2x''(t)+(1+\gamma^2)x(t)=0.$$
We seek solutions $x(t)=e^{i\nu t}$ and find that $\lambda=\nu^2$ satisfies the
quadratic equation $\lambda^2-2\lambda+1+\gamma^2=0$, so $\nu=\pm\sqrt{1\pm i
\gamma}$. Thus, the characteristic frequencies are always complex. By
decomposing $\sqrt{1+i\gamma}=a+ib$ into its real and imaginary parts we can
write the general solution as
$$x(t)=\left[(A+D)e^{-bt}+(B+C)e^{bt}\right]\cos(at)
+i\left[(A-D)e^{-bt}+(B-C)e^{bt}\right]\sin(at),$$
where $A$, $B$, $C$ and $D$ are arbitrary constants. Therefore, for any initial
conditions, the real and imaginary parts of $x(t)$ are oscillatory and growing
(or decreasing). As a consequence, the trajectories in the complex-$x$ plane
spiral outward (or inward). Hence, the classical paths are open (see
Fig.~\ref{F8}, left panel). Thus, although the quantum spectrum is partially
real, this partial reality does not give rise to closed classical trajectories.

\begin{figure}
\centering
\includegraphics[width = 115mm]{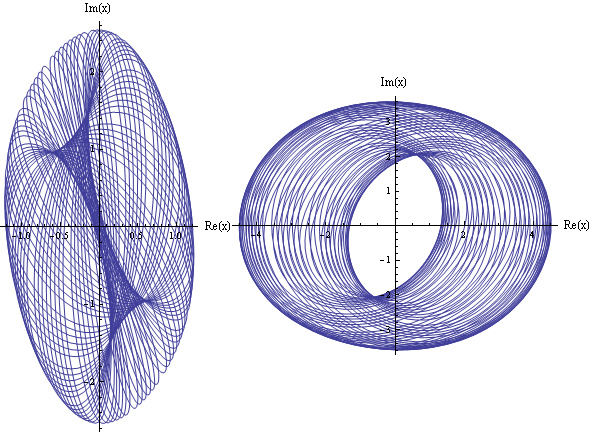}
\caption{Classical trajectory in the complex-$x$ plane for $H_2$ with the
parameter choice $\omega_x^2=3$, $\omega_y^2=1$, $\gamma=1/2$, with the
initial conditions $x(0)=-1+i$, $y(0)=2-2i$, $\dot{x}(0)=1+i/2$, $\dot{y}(0)=
3/2+i$ (left panel) and $x(0)=1+2i$, $y(0)=2+i/4$, $\dot{x}(0)=-3+i$,
$\dot{y}(0)=1/2+5i$ (right panel).}
\label{F9}
\end{figure}

More generally, if the coupled oscillators described by $H_2$ have natural
frequencies $\omega_x$ and $\omega_y$, we obtain the equation
$$x''''(t)+\left(w_x^2+w_y^2\right)x''(t)+\left(w_x^2w_y^2+\gamma^2\right)x(t)
=0.$$
Again seeking solutions $x(t)=e^{i\nu t}$, we find that $\nu^2=\half w_x^2+
\half w_y^2\pm\half\big[(\omega_x^2-\omega_y^2)^2-4\gamma^2\big]^{1/2}$. We
deduce that four real values of $\nu$ exist when $|\omega_x^2-\omega_y^2|\geq2|
\gamma|$, which is precisely the condition that guarantees a fully real spectrum
in the quantum system [see (\ref{E20})]. Thus, the transition from the broken
partial-$\cPT$-symmetric phase to the unbroken phase occurs at the same point as
for the quantum case. For the parameter choice $\omega_x^2=2$, $\omega_y^2=1$,
and $\gamma=1/2$ the classical trajectory depicted in Fig.~\ref{F8} (right
panel) spirals outward, which indicates that $\cPT$ symmetry is broken even
though the system is partially $\cPT$ symmetric. The behavior of the
trajectories in the unbroken phase is illustrated in Fig.~\ref{F9}. Here, one
sees that while the classical trajectory is not closed it is confined to a
compact region in the complex-$x$ plane. We thus observe the phase transition at
the classical level; that is, we observe the transition from spirals
(broken phase) to localized trajectories in the complex-$x$ plane (unbroken
phase), which happens at $|\omega_x^2-\omega_y^2|=2|\gamma|$.

\begin{figure}
\centering
\includegraphics[width=140mm]{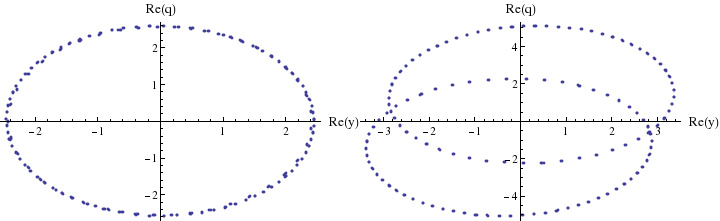}
\caption{Poincar\'e section of the classical trajectory in phase space for $H_2$
with the parameter choice $\omega_x^2=3$, $\omega_y^2=1$, $\gamma=1/2$ and
initial conditions $x(0)=1+i/2$, $y(0)=-2+2i$, $\dot{x}(0)=1+i/2$, $\dot{y}(0
)=-3/2-3i/2$ (left panel) and $x(0)=2+5i/2$, $y(0)=2-3i$, $\dot{x}(0)=1-i/2$,
$\dot{y}(0)=2-2i$ (right panel). The structure of this plot indicates that
the orbits are almost periodic.}
\label{F10}
\end{figure}

In addition to studying the trajectory in the complex-$x$ plane, one can study
the trajectories in phase space by plotting a Poincar\'e section. From the
structure of the Poincar\'e plot we conclude that the confined trajectories are
{\it almost periodic} \cite{BO}. This is illustrated in Fig.~\ref{F10}. The
appearance of open almost-periodic trajectories occurs because the number of
degrees of freedom exceeds one; if $N=1$, the classical orbits associated
with real quantum energies are closed \cite{BBH}.

\begin{figure}
\centering
\includegraphics[width=140mm]{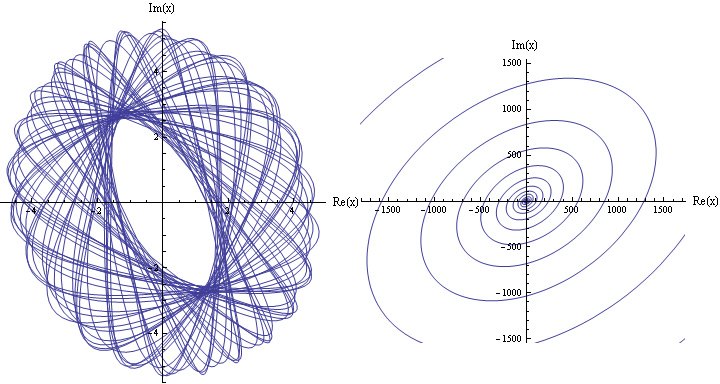}
\caption{Classical trajectory in the complex-$x$ plane for $H_3$ with the
parameter choice $\omega_x^2=1/3$, $\omega_y^2=2/3, \gamma=1/12$, $\omega_z^2=1$
(left panel) and $\omega_z^2=13/20$ (right panel) with the initial conditions
$x(0)=-2+i$, $y(0)=3-3i$, $z(0)=3+2i$, $\dot{x}(0)=-1+3i$, $\dot{y}(0)=3+2i$,
$\dot{z}(0)=-2+i$.}
\label{F11}
\end{figure}

\begin{figure}
\centering
\includegraphics[width=80mm]{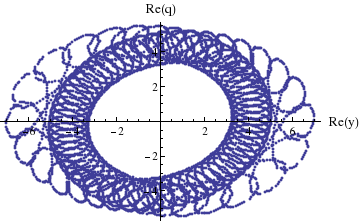}
\caption{Poincar\'e section of the classical trajectory for $H_3$ with the
parameter choice $\omega_x^2=1/3$, $\omega_y^2=2/3$, $\omega_z^2=1$, $\gamma=1/
12$ and the initial conditions $x(0)=-2+i$, $y(0)=3-3i$, $z(0)=3+2i$, $\dot{x}(
0)=-1+3i$, $\dot{y}(0)=3+2i$, $\dot{z}(0)=-2+i$. This figure indicates that
the orbit is almost periodic.}
\label{F12}
\end{figure}

A similar analysis can be done for the three-coupled-oscillator Hamiltonian
$H_3$ with general natural frequencies. The classical equations of motion are
$$x''(t)+\omega_x^2x(t)+i\gamma y(t)=0,\quad y''(t)+\omega_y^2y(t)+i\gamma(x(t)+
z(t))=0, \quad z''(t)+\omega_z^2z(t)+i\gamma y(t)=0.$$
Seeking solutions of the form $x(t)=Ae^{i\nu t}$, $y(t)=Be^{i\nu t}$ and $z(t)=C
e^{i\nu t}$, we obtain a cubic equation for $\lambda=\nu^2$:
$$\lambda^3-\left(\omega_x^2+\omega_y^2+\omega_z^2\right)\lambda^2+
\left(\omega_x^2\omega_y^2+\omega_x^2\omega_z^2+\omega_y^2\omega_z^2+2\gamma^2
\right)\lambda-\omega_x^2\omega_y^2\omega_z^2-\left(\omega_x^2+\omega_z^2\right)
\gamma^2=0.$$
The characteristic frequencies are real if and only if the corresponding quantum
system is in an unbroken $\cPT$-symmetric phase (all eigenvalues are real); the
criteria for real eigenvalues is given in Sec.~\ref{s4}. To illustrate, recall
that in Sec.~\ref{s4} the parameter choice $\omega_x^2=1/3$, $\omega_y^2=2/3$,
$\omega_z^2=1$, $\gamma=1/12$ lies in the unbroken phase and gives a real energy
spectrum. Figure~\ref{F11} shows that for this parameter choice, the classical
trajectory is confined to a compact region in the complex-$x$ plane, whereas for
the choice $\omega_x^2=1/3$, $\omega_y^2=2/3$, $\omega_z^2=13/20$, and $\gamma=1
/12$, for which the quantum symmetry is broken, the classical trajectory spirals
outward to infinity. A Poincar\'e section is given in Fig.~\ref{F12}
for the parameter choice of Fig.~\ref{F11}, left panel.

\section{Brief concluding remarks}
\label{s6}
In this paper we have examined systems of $N$ linearly coupled oscillators that
are partially $\cP\cT$-symmetric. In the quantum-mechanical analysis we have
found that the ground state of each of these systems is always real. We have
shown that the entire spectrum may in fact be completely real depending on the
values of the natural frequencies $\omega_x$, $\omega_y$, $\omega_z$, $\dots$
and their relation to the coupling strength $\gamma$. This happens even though
the system is only partially $\cPT$ symmetric. We have studied this in
detail for systems of two and three coupled oscillators. A phase transition
point exists beyond which the energy spectrum is only partially real. 

For the two and three classical oscillator systems, we find a phase transition
at exactly the same point as the quantum-mechanical oscillator systems. When the
eigenvalues of the quantum system are all real, the classical trajectories
are confined and almost periodic, but when the quantum eigenvalues are
partly real and partly complex, the corresponding classical system always has
open trajectories that spiral out to infinity.

\acknowledgments
CMB thanks the Heidelberg Graduate School of Fundamental Physics for its
hospitality.


\begin{thebibliography}{17}

\bibitem{r1} J.~Schindler, A.~Li, M.~C.~Zheng, F.~M.~Ellis, and T.~Kottos,
Phys.~Rev.~A {\bf 84}, 040101 (2011).

\bibitem{r2} C.~M.~Bender, B.~Berntson, D.~Parker, and E.~Samuel,
Am.~J.~Phys. {\bf 81}, 173 (2013).

\bibitem{r3} C.~M.~Bender, M.~Gianfreda, S.~K.~\"Ozdemir, B.~Peng, and
L.~Yang, Phys.~Rev.~A {\bf 88}, 062111 (2013).

\bibitem{r4} J.~Cuevas, P.~G.~Kevrekidis, A.~Saxena, and A.~Khare,
Phys.~Rev.~A {\bf 88}, 032108 (2013).

\bibitem{r5} C.~M.~Bender, M.~Gianfreda, and S.~P.~Klevansky, Phys.~Rev.~A
{\bf 90}, 022114 (2014).

\bibitem{r6} B.~Peng, S.~K.~\"Ozdemir, F.~Lei, F.~Monifi, M.~Gianfreda,
G.~L.~Long, S.~Fan, F.~Nori, C.~M.~Bender, L.~Yang, Nat.~Phys.~{\bf 10}, 394
(2014).

\bibitem{r7} I.~L.~Aleiner, B.~L.~Altshuler, and Y.~G.~Rubo, Phys.~Rev.~B
{\bf 85}, 121301 (2012).

\bibitem{BO} C.~M.~Bender and S.~A.~Orszag, {\it Advanced Mathematical
Methods for Scientists and Engineers} (McGraw-Hill, New York, 1978).

\bibitem{BBH} C.~M.~Bender, D.~C.~Brody, and D.~W.~Hook,
J.~Phys.~A: Math.~Theor.~{\bf 41}, 352003 (2008).

\end{thebibliography}
\end{document}